\documentclass[aps,preprintnumbers, superscriptaddress, amsmath, showpacs, showkeys]{revtex4-1}
\usepackage{epsfig}
\usepackage{psfrag}

\usepackage{graphicx}
\usepackage{dcolumn}
\usepackage{bm}

\begin{document}

\title{Magnetoelectric Effect in Strongly Magnetized Color Superconductivity}

\author{Bo  Feng,  Efrain J. Ferrer  and   Vivian de la Incera\\\textit{Department of Physics, University of Texas at El Paso, El Paso, Texas
79968, USA}}





\date{\today}

\begin{abstract}
The effect of a strong magnetic field on the electric polarization of a three-flavor color superconducting medium is investigated. We found that the electric susceptibility of this strongly magnetized medium is highly anisotropic. In the direction transverse to the applied magnetic field the susceptibility reduces to that of the vacuum, while in the longitudinal direction it depends on the magnetic field and decreases with it. The nature of this behavior is associated with the field's dependence of the Cooper pairs' coherence length, which plays the role of the electric dipole length. The field's dependence of the electric polarization is interpreted as the realization of the magnetoelectric effect in cold-dense quark matter.
\end{abstract}

\pacs{26.30.+k, 91.65.Dt, 98.80.Ft}
\keywords{Color Superconductivity, Magnetized QCD, Magnetoelectricity, Magnetars}
\maketitle

\section{Introduction}
By now, it has been well established that color superconductivity (CS) is the most favored state of nuclear matter at low temperatures and extremely high densities \cite{CS}. The best candidates for the realization of CS are the core of compact stars, which are, besides black holes, the most dense known objects in nature. Compact stars, on the other hand, are typically strongly magnetized objects. Specifically, the so called magnetars have surface magnetic fields as large as $10^{14}-10^{15}$ G \cite{Magnetars}. Because the stellar medium has very high electric conductivity, the magnetic flux should be conserved there. It is then reasonable to assume that in the region of the star with the largest matter density, i.e. the core, the magnetic field strength is even larger than in the surface. However, the interior magnetic fields of neutron stars are not directly accessible to observation, so their possible values can only be estimated with the help of heuristic methods. Estimates based on macroscopic and microscopic analysis, for nuclear \cite{virial}, and quark matter, considering both gravitationally bound and self-bound stars \cite{EoS-H}, have led to maximum fields within the range of $10^{18}$ and $10^{20}$ G, depending respectively if the inner medium is formed by neutrons \cite{virial}, or quarks \cite{EoS-H}.

Let us recall that even though the original electromagnetic $U(1)_{em}$ symmetry is broken by the formation of quark Cooper pairs in the Color-Flavor-Locked (CFL) phase \cite{CFL} of CS, a residual $\widetilde{U}(1)$ symmetry still remains. The massless gauge field associated with this symmetry is given by the linear combination of the conventional photon field and the $8^{th}$ gluon field \cite{CFL, ABR}, ${\tilde A}_\mu=\cos\theta A_\mu-\sin\theta G^8_\mu$. The field $\tilde A_{\mu}$ plays the role of an in-medium or rotated electromagnetic field. A magnetic field associated with $\tilde A_{\mu}$ can penetrate the CS without being subject to the Meissner effect, since the color condensate is neutral with respect to the corresponding rotated charge. Actually, the penetrating field in the CFL superconductor is mostly formed by the original photon with only a small admixture of the 8th gluon since the mixing angle, $\cos^{-1}\theta=g/\sqrt{e^2/3+g^2}$, is sufficiently small. A similar residual electromagnetic group also remains in the 2SC phase of CS \cite{2SC}.

The unbroken $\widetilde U(1)$ symmetry corresponding to the long-range rotated photon in CFL is generated by ${\tilde Q}=Q\times 1+1\times T_8/\sqrt{3}$, where $Q$ is the conventional electromagnetic charge operator of quarks and $T_8$ is the 8th Gell-Mann matrix. Using the matrix representations, $Q=diag(-1/3,-1/3,2/3)$ for $(s,d,u)$ flavors, and $T_8=diag(-1/\sqrt{3},-1/\sqrt{3},2/\sqrt{3})$ for $(b,g,r)$ colors, the ${\tilde Q}$ charges (in units of ${\tilde e}=e\cos\theta$) of different quarks are
\begin{equation}\label{table}
\begin{tabular}{ccccccccc}
\hline
\textrm{$s_b$}&
\textrm{$s_g$}&
\textrm{$s_r$}&
\textrm{$d_b$}&
\textrm{$d_g$}&
\textrm{$d_r$}&
\textrm{$u_b$}&
\textrm{$u_g$}&
\textrm{$u_r$}\\
\colrule
0 & 0 & $-1$ & 0 & 0 & $-1$ & $+1$ & $+1$ & 0\\
\hline
\end{tabular}
\end{equation}

The less symmetric realization of the CFL pairing that occurs in the presence of a magnetic field, is known as the magnetic-CFL (MCFL) phase \cite{MCFL}. The MCFL phase has similarities, but also important differences with the CFL phase \cite{MCFL,MCFLoscillation,phases}.  For example, the ground state has different symmetry and is characterized by two antisymmetric gaps $\Delta$ and $\Delta_B$, instead of just one, as in the regular CFL case. Recently, it was also found that in the MCFL phase a new condensate $\Delta_M$, associated with the magnetic moment of the Cooper pairs, is also present \cite{SpinoneCFL}. The extra condensate appears because in this phase the Cooper pairs formed by charged quarks have nonzero magnetic moment, since the quarks in the pair not only have opposite charge but also opposite spin. The magnetic moment of this type of pairs leads to a nonzero net magnetic moment for the system, which in turn is reflected in the existence of a new condensate $\Delta_M$, in addition to the gaps $\Delta$ and $\Delta_{B}$ \cite{SpinoneCFL}. This new condensate is a spin-one order parameter and since it is a direct consequence of the external magnetic field, its magnitude becomes comparable to the other energy gaps only at strong field values. The $\Delta_M$ condensate of the MCFL phase shares a few similarities with the dynamical generation of an anomalous magnetic moment recently found in massless QED \cite{magmoment}.

It is well known that the phenomenon of CS shares many characteristics of condensed matter systems \cite{QCDcondense}. In this paper, we discovered a new feature of CS that has its counterpart in magnetically ordered materials and has been known in the context of condensed matter for many years. It is the so called magnetoelectric (ME) effect, which establishes a relation between the electric and magnetic properties of certain materials. In general, it states that the electric polarization of such materials may depend on an applied magnetic field and/or that the magnetization may depend on an applied electric field. The first observations of magnetoelectricity took place when a moving dielectric was found to become polarized when placed in a magnetic field \cite{early-ME}. In 1894, Pierre Curie \cite{Curie} was the first in pointing out the possibility of an intrinsic ME effect for certain (non-moving) crystals on the basis of symmetry considerations. But it took many decades to be understood and proposed by Landau and Lifshitz \cite{L-L} that the linear ME effect is only allowed in time-asymmetric systems. Recently the ME effect regained new interest in condensed matter thanks to new advancements in material science and with the development of the so-called multiferroic materials for which the ME effect is significant for practical applications \cite{Revival}.

In this paper we will show that the ME effect is also realized in a highly magnetized CS medium like the MCFL phase of color superconductivity. In particular, we will find how the electric susceptibility of this medium depends on an applied strong magnetic field.

\section{Magnetoelectric effect in color superconductivity}\label{sec:level1}

As already pointed out in the Introduction, the ME phenomenon is associated with the variation of the electric polarization with a magnetic field or the variation of the magnetization with an electric field, although this last case will not be considered here. In this section we will discuss, on general grounds, how the ME effect can take place in a strongly magnetized, highly dense three-flavor color superconductor characterized by the MCFL phase.

Let us start by discussing the ME effect at weak fields. At weak fields this effect can be studied by taking into account the expansion of the system's free energy in powers of the electric $\bf \widetilde{E}$ and magnetic $\bf \widetilde{B}$ fields
\begin{equation}\label{free-energy}
F({\bf \widetilde{E}},{\bf \widetilde{B}})=F_0-\alpha_i\widetilde{E}_i-\beta_i\widetilde{B}_i-\gamma_{ij}\widetilde{E}_i\widetilde{B}_j-\eta_{ij}\widetilde{E}_i\widetilde{E}_j-\tau_{ij}
\widetilde{B}_i\widetilde{B}_j-\kappa_{ijk}\widetilde{E}_i\widetilde{E}_j\widetilde{B}_k-\lambda_{ijk}\widetilde{E}_i\widetilde{B}_j\widetilde{B}_k-\sigma_{ijkl}\widetilde{E}_i\widetilde{E}_j\widetilde{B}_k\widetilde{B}_l-\ldots
\end{equation}
In this weak-field expansion the coefficients $\alpha_i, \gamma_{ij}$ etc., which are the susceptibility tensors, can be found from the infinite set of one-loop polarization operator diagrams with external legs of the in-medium photon field $\widetilde{A}_{\mu}$ and internal lines of the full CFL quark propagator of the rotated charged quarks. Hence, these coefficients can only depend on the baryonic chemical potential, the temperature and the CFL gap. From (\ref{free-energy}), the electric polarization can be found as
\begin{equation} \label{Polarization}
P_i=-\frac{\partial F}{\partial \widetilde{E}_i}=\alpha_i+\gamma_{ij}\widetilde{B}_j+2\eta_{ij}\widetilde{E}_j+2\kappa_{ijk}\widetilde{E}_j\widetilde{B}_k+\lambda_{ijk}\widetilde{B}_j\widetilde{B}_k+2\sigma_{ijkl}\widetilde{E}_j\widetilde{B}_k\widetilde{B}_l+\ldots
\end{equation}

If the tensor $\gamma$ is different from zero the system exhibits the linear ME effect. From the free energy (\ref{free-energy}), we see that the linear ME effect can only exist if the time-reversal and parity symmetries are broken in the medium. In the CFL phase, the time-reversal symmetry is broken by the CFL gap \cite{SpinoneCFL}, but parity is preserved. Thus, the linear ME effect cannot be present in this medium. The behavior under a time-reversal transformation underscores an important difference between the CFL color superconductivity and the conventional, electric superconductivity. While the CFL color superconductor is not invariant under time-reversal symmetry, the conventional superconductor is, since in the conventional superconductor the Cooper pairs are usually formed by time-reversed one-particle states \cite{Anderson}. In the conventional superconductor the violation of the T-invariance occurs only via some external perturbation which can lead in turn to pair breaking and to the so-called gapless superconductivity \cite{Abrikosov}.

Higher-order ME terms are parameterized by the tensors $\kappa$, and $\lambda$. As it happens with $\gamma$, the coefficient $\lambda \neq 0$ is forbidden because it requires parity violation. On the other hand, although a $\kappa \neq 0$ term only requires time-reversal violation, to form a third-rank tensor independent of the momentum and parity invariant, the medium would need to have an extra spatial vector structure. However, the only tensor structures available to form such a third-rank tensor in the CFL phase are the metric tensor $g_{\mu \nu}$  and the medium fourth velocity $u_\mu$, which in the rest frame is a temporal vector $u_\mu=(1,0,0,0)$, so the coefficient $\kappa$ should be zero too. Hence, we do not expect any ME effect associated with the lower terms in the weak-field expansion of the free energy (\ref{Polarization}).

At strong magnetic fields, the situation is quite different. In this case the expansion of the free energy can only be done in powers of a weak electric field, and the coefficients of each term can be found from the corresponding one-loop polarization operators, which now depend on the strong magnetic field in the MCFL phase. The free energy expansion in this case takes the form
\begin{equation}\label{free-energy-strong}
F'({\bf \widetilde{E}},{\bf \widetilde{B}})=F'_0(\widetilde{B})-\alpha'_i\widetilde{E}_i-\eta'_{ij}\widetilde{E}_i\widetilde{E}_j-\ldots
\end{equation}
The tensors $\alpha'$ and $\eta'$ can depend now on the baryonic chemical potential, temperature, magnetic field and gaps of the MCFL phase \cite{SpinoneCFL}. They can be found respectively by calculating the tadpole and the second rank polarization operator tensor of the MCFL phase in the strong field limit. An $\alpha'\neq 0$ would indicate that the MCFL medium behaves as a ferroelectric material \cite{ferroelectricity}, but this is not the case because this phase is parity symmetric \cite{SpinoneCFL}, hence $\alpha' = 0$. The tensor $\eta'$, nevertheless, is not forbidden by any symmetry argument. If it is different from zero, $\eta'$ would characterize the lowest order of the system dielectric response. More important, if $\eta'$ results to be dependent on the magnetic field, this would imply that the electric polarization $P=\eta' E$ depends on the magnetic field through $\eta'$, hence the MCFL phase would exhibit the ME effect.

The main goal of this paper is to find the electric susceptibility $\eta'$ in the strong-magnetic-field limit of the MCFL phase \cite{SpinoneCFL}. Taking into account that
\begin{equation}\label{Strong-B-Suscept}
   F'({\bf \widetilde{E}},{\bf \widetilde{B}})-F'_0({\bf\widetilde{B}})\sim\frac{1}{V}\int \widetilde{A}_0(x_3)\Pi_{00}(x_3-x_3')\widetilde{A}_0(x_3')dx_3dx_3'=-\eta' \widetilde{E}^2,
\end{equation}
our task can be reduced to the calculation of the zero-zero component of the one-loop polarization operator at strong magnetic field in the infrared limit, $\Pi_{00}(p_0=0, p\rightarrow 0)$.

\section{The one-loop polarization operator of the rotated photon at $\widetilde{B} \neq 0$}

In coordinate space the one-loop polarization operator for the rotated photon in the MCFL phase reads
\begin{equation}
\Pi^{\mu\nu}(x,y)=\frac{{\tilde e}^2}{2}\sum_{{\tilde Q}=\pm}Tr\left[\Gamma^\mu{\cal S}_{(\tilde Q)}(x,y)\Gamma^\nu{\cal S}_{(\tilde Q)}(y,x)\right]
\label{selfenergy}
\end{equation}
where ${\cal S}_{(\tilde Q)}(x,y)$ and $\Gamma^\mu$, defined in the Nambu-Gorkov space, are the full quark propagator
\begin{equation}\label{full-propagator}
S_{(\tilde Q)}(x,x^\prime)=\int\hspace{-0.5cm}\sum\frac{d^4p}{(2\pi)^4}E^{l{(\tilde Q)}}_p(x)\Pi(l){\tilde S}_{(\tilde Q)}^l({\bar p}^{(\tilde Q)}){\bar E}^{l{(\tilde Q)}}_p(x^\prime)
\end{equation}
with
\begin{equation}\label{full-prog-matrix}
{\tilde S}_{(\tilde Q)}^l({\bar p}^{(\tilde Q)})=\left(
\begin{array}{cc}
{G^+_{(\tilde Q)}}^{l}(\bar p^{(\tilde Q)}) & {\Xi^-_{(\tilde Q)}}^{l}(\bar p^{(\tilde Q)})\\
{\Xi^+_{(\tilde Q)}}^{l}(\bar p^{(\tilde Q)}) & {G^-_{(\tilde Q)}}
^{l}(\bar p^{(\tilde Q)})
\end{array}
\right)
\end{equation}
and vertex
\begin{equation}
\Gamma^\mu=\left(
\begin{array}{cc}
{\tilde Q}\gamma^\mu & 0\\
0 & {\tilde Q}\gamma^\mu
\end{array}
\right)
\end{equation}
respectively.
In (\ref{full-propagator}), the notation $\int\hspace{-0.38cm}\sum\frac{d^4p}{(2\pi)^4}\equiv\sum^\infty_{l=0}\int\frac{dp_0dp_2dp_3}{(2\pi)^4}$ is understood. The projector $\Pi(l)\equiv \Delta(\textrm{sgn}(\tilde Q\widetilde{B}))\delta_{l0}+I(1-\delta_{l0})$ addresses the difference between the spin degeneracy of nonzero
and zero Landau levels,  and the spin projectors  $\Delta(\pm)=(1+i\gamma^1\gamma^2)/2$ correspond respectively to spin up $(+)$ and down $(-)$ if $\textrm{sgn}(\tilde Q\widetilde{B})>0$, or viceversa if  $\textrm{sgn}(\tilde Q\widetilde{B})<0$. Notice that in (\ref{selfenergy}) only quarks with nonzero rotated charge contribute to the rotated-photon self-energy (see that in the sum $\widetilde{Q}=\pm$ only).

In the strong field approximation, we only need to consider the matrix elements of (\ref{full-prog-matrix}) in the lowest Landau level (LLL) $l=0$, where
\begin{equation}\label{G-LLL}
{G^\pm_{(+)}}^{l=0}(p^\parallel)={G^\pm_{(-)}}^{l=0}(p^\parallel)=\sum_{e=\pm}\frac{p_0\mp(\mu-ep_3)}{p_0^2-[\epsilon_{p_3}^e]^2}\Lambda_{{\bf p}_3}^{\pm e}\gamma_0
\end{equation}
and
\begin{equation}\label{Xi-LLL}
{\Xi^\pm_{(+)}}^{l=0}(p^\parallel)={\Xi^\pm_{(-)}}^{l=0}(p^\parallel)=\pm\sum_{e=\pm}\frac{\Delta_{0}}{p_0^2-[\epsilon_{p_3}^e]^2}\gamma_5\Lambda_{{\bf p}_3}^{\mp e}
\end{equation}
The effective gap of the LLL modes is given by the combination $\Delta_{0}=\Delta_M-\Delta_B$ \cite{SpinoneCFL}; $\Lambda_{{\bf p}_3}^e=(1+e\gamma_0\gamma^3{\hat{\bf p}}_3)/2$ are the projectors onto states of positive $(e=+)$ or negative $(e=-)$ energy with $\widehat{\textbf{p}}_3=p_3/|p_3|$, and $\epsilon_{p_3}^e\equiv\sqrt{(\mu-ep_3)^2+\Delta_{0}^2}$ are the quasiparticle energies. Notice that, in the LLL, the terms (\ref{G-LLL}) and (\ref{Xi-LLL}) only depend on the parallel momenta. The variation of $\Delta_B$ and $\Delta_M$ with $B$ was found in \cite{SpinoneCFL}. In transforming the full propagator to momentum space in (\ref{full-propagator}), we used the so-called Ritus's method \cite{Ritus}. In this approach, the transformation to momentum space is carried out by the eigenfunctions ${\bf E}^{l(\pm)}_p(x)=E^{l(\pm)}_p\Delta(\pm)+E^{l-1(\pm)}_p\Delta(\mp)$ of the asymptotic states of the charged fermions in a uniform magnetic field, and $E^{l(\pm)}_p(x)={\cal N}_le^{-i(p_0x^0+p_2x^2+p_3x^3)}D_l(\rho_{(\pm)})$ are the corresponding eigenfunctions with normalization constant ${\cal N}_l=(4\pi {\tilde e}{\tilde B})^{1/4}/\sqrt{l!}$. $D_l(\rho_{(\pm)})$ denotes the parabolic cylinder functions of argument $\rho_{(\pm)}=\sqrt{2{\tilde e}{\tilde B}}(x_1\pm p_2/{\tilde e}{\tilde B})$, and index given by Landau level $l=0,1,2,...$

To find the polarization operator in momentum space we assume translational invariance for the photon self-energy,
\begin{equation}\label{Pi-transf}
(2\pi)^4\delta^{(4)}(p-p^\prime)\Pi^{\mu\nu}(p)=\int d^4xd^4x^\prime e^{-i(p\cdot x-p^\prime\cdot x^\prime)}\Pi^{\mu\nu}(x,x^\prime)
\end{equation}
Now, substituting with (\ref{full-propagator}) into (\ref{Pi-transf}) the polarization operator in momentum space becomes
\begin{align}
\nonumber(2\pi)^4\delta^{(4)}(p-p^\prime)\Pi^{\mu\nu}(p)=&\frac{{\tilde e}^2}{2}\sum_{\tilde Q}{\tilde Q}^2\int d^4xd^4x^\prime\int\hspace{-0.5cm}\sum\frac{d^4k}{(2\pi)^4}\int\hspace{-0.5cm}\sum\frac{d^4q}{(2\pi)^4}e^{-i(p\cdot x-p^\prime\cdot x^\prime)}\\
&\times{\rm Tr}\left[\gamma^\mu {\bf E}^{l{(\tilde Q)}}_k(x)\Pi(l){\tilde S}_{{(\tilde Q)}}^l({\bar k}^{(\tilde Q)}){\bar{\bf E}}^{l{(\tilde Q)}}_k(x^\prime)\gamma^\nu {\bf E}^{m{(\tilde Q)}}_q(x^\prime)\Pi(m){\tilde S}_{(\tilde Q)}^m({\bar q}^{(\tilde Q)}){\bar{\bf E}}^{m{(\tilde Q)}}_q(x)\right]
\label{selfenergyinmomentum}
\end{align}
At this point we can use the integral formulas \cite{Ng}
\begin{align}
\nonumber\int d^4x {\bar{\bf E}}^{m({\tilde Q})}_q(x)\gamma^\mu{\bf E}^{l({\tilde Q})}_k(x)e^{-ip\cdot x}=&(2\pi)^4\delta^{(3)}(q+p-k)e^{-ip_1(k_2+q_2)/{2{\tilde e}{\tilde B}}}e^{-{p}^2_\perp/2}\\
&\times\sum_{\sigma, \sigma^\prime}\frac{1}{\sqrt{n!n^\prime !}}e^{i\textrm{sgn}({\tilde e}{\tilde B})(n-n^\prime)\varphi}J_{nn^\prime}({\hat p}_\perp)\Delta(\sigma)\gamma^\mu\Delta(\sigma^\prime)
\label{integralone}
\end{align}
with $n\equiv n(m,\sigma)$ and $n^\prime\equiv n(l,\sigma^\prime)$ defined by $n(l,\sigma)=l+\textrm{sgn}({\tilde e}{\tilde B})\frac{\sigma}{2}-\frac{1}{2}$. The transverse momentum and the polar angle are defined in terms of the dimensionless variables ${\hat p}_\mu\equiv p_\mu/\sqrt{2|{\tilde e}{\tilde B}|}$ as ${\hat p}_\perp\equiv \sqrt{{\hat p}^2_1+{\hat p}^2_2}$ and $\varphi\equiv \arctan({\hat p}_2/{\hat p}_1)$, respectively; the delta function $\delta^{(3)}(q+p-k)\equiv \delta(q_0+p_0-k_0)\delta(q_2+p_2-k_2)\delta(q_3+p_3-k_3)$ and
\begin{equation}
J_{nn^\prime}({\hat p}_\perp)=\sum_{m=0}^{\textrm{min}(nn^\prime)}\frac{n!n^\prime !}{m!(n-m)!(n^\prime-m)!}\left[i\textrm{sgn}({\tilde e}{\tilde B}){\hat p}_\perp\right]^{n+n^\prime-2m}
\end{equation}
The presence of the delta functions in (\ref{integralone}) facilitates the integrations over $k_0, k_2$ and $k_3$ in (\ref{selfenergyinmomentum}) yielding an overall $\delta^{(3)}(p-p^\prime)=\delta(p_0-p_0^\prime)\delta(p_2-p_2^\prime)\delta(p_3-p_3^\prime)$ and the integral over $q_2$ gives rise to $\delta(p_1-p_1^\prime)$, which combined with the previous deltas matches the delta function product on the LHS of (\ref{selfenergyinmomentum}). Thus, we have
\begin{align}\label{pol-op-1}
\nonumber\Pi^{\mu\nu}(p)=&\frac{{\tilde e}^2}{2}({\tilde e}{\tilde B})\sum_{\tilde Q}\sum_l\sum_{[\sigma]}{\tilde Q}^2\int\hspace{-0.5cm}\sum\frac{d^3q}{(2\pi)^3}\frac{e^{i\textrm{sgn}({\tilde e}{\tilde B})(n-n^\prime+{\bar n}-{\bar n}^\prime)\varphi}}{\sqrt{n!n^\prime!{\bar n}!{\bar n}^\prime!}}e^{-{\hat p}^2_\perp}J_{nn^\prime}({\hat p}_\perp)J_{{\bar n}{\bar n}^\prime}({\hat p}_\perp)\\
&\times{\rm Tr}\left[\Delta(\sigma)\gamma^\mu\Delta(\sigma^\prime)\Pi(l){\tilde S}_{(\tilde Q)}^l({\bar p}^{(\tilde Q)}-{\bar q}^{(\tilde Q)})\Delta({\bar\sigma})\gamma^\nu\Delta({\bar\sigma}^\prime)\Pi(m){\tilde S}_{(\tilde Q)}^m({\bar q}^{(\tilde Q)})\right]
\end{align}
with $\int\hspace{-0.4cm}\sum\frac{d^3q}{(2\pi)^3}=\sum_m\int dq_0dq_3$ and $[\sigma]$ meaning summing over $\sigma, \sigma^\prime, {\bar\sigma}$ and ${\bar\sigma}^\prime$. Because of the factor $e^{-{\hat p}^2_\perp}$ in the integrand, contributions from large values of ${\hat p}_\perp$ are suppressed. Thus, by keeping only the terms with the smallest power of ${\hat p}_\perp$ in $J_{nn^\prime}({\hat p}_\perp)$ we have $J_{nn^\prime}({\hat p}_\perp)\rightarrow n!\delta_{n,n^\prime}$. Then the polarization operator reads
\begin{align}\label{pol-op-2}
\nonumber\Pi^{\mu\nu}(p)=&\frac{{\tilde e}^2}{2}({\tilde e}{\tilde B})\sum_{\tilde Q}\sum_l\sum_{\sigma}{\tilde Q}^2\int\hspace{-0.5cm}\sum\frac{d^3q}{(2\pi)^3}e^{-{\hat p}^2_\perp}\delta_{n,n^\prime}\delta_{{\bar n},{\bar n}^\prime}\\
&\times{\rm Tr}\left[\Delta(\sigma)\gamma^\mu\Delta(\sigma^\prime)\Pi(l){\tilde S}_{(\tilde Q)}^l({\bar p}^{(\tilde Q)}-{\bar q}^{(\tilde Q)})\Delta({\bar\sigma})\gamma^\nu\Delta({\bar\sigma}^\prime)\Pi(m){\tilde S}_{(\tilde Q)}^m({\bar q}^{(\tilde Q)})\right]
\end{align}
The polarization operator at finite temperature $T$ can be readily found by replacing the integral in $q_0$ by a sum in the Matsubara's frequencies $q_0=i(2k+1)\pi/\beta$, $k=0,\pm 1, \pm 2,...$. Taking into account that $\delta_{n,n^\prime}=\delta_{m,l}\delta_{\sigma\sigma^\prime}+\delta_{m+\sigma,l}\delta_{-\sigma,\sigma^\prime}$ and $\delta_{{\bar n},{\bar n}^\prime}=\delta_{m,l}\delta_{{\bar\sigma},{\bar\sigma}^\prime}+\delta_{m+{\bar\sigma}^\prime,l}\delta_{-{\bar\sigma}^\prime,{\bar\sigma}}$,
we can sum in $[\sigma]$ and $m$, to obtain
\begin{equation}
\Pi_{\mu\nu}(p^{\parallel})=
\tilde e^2({\tilde e}{\tilde B})T\sum_{q_0}\int\frac{dq_3}{(2\pi)^2}{\rm Tr}\left[\Delta(+)\gamma^{\parallel}_\mu{\tilde S}^{0}_{(+)}({p^\parallel-q^\parallel})\Delta(+)\gamma_\nu^\parallel{\tilde S}^{0}_{(+)}({q}^\parallel)\right]
\label{polarizationtensor}
\end{equation}
Therefore, in the LLL approximation $\Pi_{\mu\nu}$ only has longitudinal components ($\mu, \nu =0,3$). Taking the traces in Nambu-Gorkov, color-flavor, and Dirac spaces, and performing the Matsubara sum we obtain
\begin{align}
\nonumber\Pi_{00}(p^\parallel)=&-{\tilde e}^2({\tilde e}{\tilde B})\int_{-\infty}^\infty \frac{dq_3}{(2\pi)^2}\sum_{e_1,e_2=\pm}(1+e_1e_2 {\hat{\bf k}}_3{\hat{\bf q}}_3)\\
\nonumber&\times\left[(\frac{1}{p_0+\epsilon_1+\epsilon_2}-\frac{1}{p_0-\epsilon_1-\epsilon_2})(1-N_1-N_2)\frac{\epsilon_1\epsilon_2-\xi_1\xi_2-\Delta^2}{2\epsilon_1\epsilon_2}\right.\\
&+\left.(\frac{1}{p_0-\epsilon_1+\epsilon_2}-\frac{1}{p_0+\epsilon_1-\epsilon_2})(N_1-N_2)\frac{\epsilon_1\epsilon_2+\xi_1\xi_2+\Delta^2}{2\epsilon_1\epsilon_2}\right]
\label{00component}
\end{align}
where $\epsilon^{e_1}_{p_{3}}=\epsilon_{1}$, $\epsilon^{e_2}_{k_{3}}=\epsilon_{2}$, $N_{1}=1/(1+e^{\beta\epsilon_1})$, $N_{2}=1/(1+e^{\beta\epsilon_2})$, $\xi_{1}=e_1q_3-\mu$ and $\xi_{2}=e_2k_3-\mu$ with $k_3=p_3-q_3$.

\section{Anisotropic Electric susceptibility}

Let us take now the static limit $p_0=0$, $p_3\rightarrow 0$ (with $p_3\ll \Delta_0$) of Eq. (\ref{00component}). In this limit, only particle-particle ($e_1=e_2=+1$) or antiparticle-antiparticle ($e_1=e_2=-1$) excitations contribute to the sum in (\ref{00component}), and moreover for $e_1=e_2=+1$
\begin{equation}
\frac{\epsilon_1\epsilon_2-\xi_1\xi_2-\Delta_{0}^2}{(\epsilon_1+\epsilon_2)\epsilon_1\epsilon_2}\cong\frac{p_3^2}{4E^2(q_3)}\frac{d^2E(q_3)}{dq_3^2},
\end{equation}
where $E(q_3)=\sqrt{(q_3-\mu)^2+\Delta_{0}^2}$ and all terms odd in $q_3-\mu$ were neglected. Then, in the zero temperature limit we obtain
\begin{equation}\label{Pi-zoro-Temp}
\Pi_{00}(p_0=0,p_3\rightarrow 0)\cong-2{\tilde e}^2({\tilde e}{\tilde B})\int_{-\infty}^\infty\frac{dq_3}{(2\pi)^2}\frac{p_3^2}{4E^2(q_3)}\frac{d^2E(q_3)}{dq_3^2}
\cong-\frac{{\tilde e}^2({\tilde e}{\tilde B})p_3^2}{6\pi^2\Delta_{0}^2}
\end{equation}

Because $\Pi_{00}(p_0=0,p_3\rightarrow 0)$ has no constant contribution, one immediately sees that there is no Debye screening in the strong-field region of the MCFL phase, as there was none either at zero-field \cite{Chargedgluon, Rischke}. It is worth noticing that the finite density, which usually leads to Debye screening in systems of free fermions, fails to produce the same effect in the MCFL color superconductor because all the quarks pair to form electrically neutral Cooper pairs. Despite the lack of Debye screening, the CS medium exhibits an anisotropic electric polarization due to its different dielectric behavior in the directions parallel and transverse to the magnetic field. As can be inferred from Eq. (\ref{Pi-zoro-Temp}), in the strong-field region the electric susceptibility of the MCFL superconductor is zero in the directions transverse to the magnetic field, but equals $\frac{{\tilde e}^2({\tilde e}{\tilde B})}{6\pi^2\Delta_{0}^2}$  in the parallel direction. The anisotropic susceptibility gives rise to a dielectric tensor with different components in the transverse and parallel directions.

From the above discussion, it is easy to understand that the effective action of the $\widetilde U(1)$ field in the strong field region can be separated in two terms
\begin{equation}
\widetilde{S}_{eff}=\int d^4x [\frac{\epsilon_{\|}}{2}{\bf \widetilde{E}_{\|}}\cdot{\bf \widetilde{E}_{\|}}+\frac{\epsilon_{\bot}}{2}{\bf \widetilde{E}_{\bot}}\cdot{\bf \widetilde{E}_{\bot}}],
\label{effectivelagrangian}
\end{equation}
Since the quadratic term in the effective $\tilde U(1)$ Lagrangian is given by $\widetilde{A}^\mu(-p)[D^{-1}_{\mu\nu}(p)+\Pi_{\mu\nu}(p)]\widetilde{A}^\nu(p)$, with $D^{-1}$ the bare rotated photon propagator, it is straightforward to find the transverse and parallel components of the dielectric tensor respectively,
\begin{equation}
\epsilon^{\bot}=1, \quad \qquad \epsilon_{\|}=1+\chi_{MCFL}^{\|}=1+\frac{{2\tilde \alpha}|{\tilde e}{\tilde B}|}{3\pi\Delta_{0}^2}
\label{transverse-epsilon}
\end{equation}
where $\widetilde{\alpha}$ is the fine-structure constant of the rotated electromagnetism.

Since (\ref{Pi-zoro-Temp}) is obtained in the strong-magnetic-field limit ${\tilde e}{\tilde B}\gg \Delta_0^2$, the parallel dielectric constant of the medium is substantially large. Thus, the Coulomb interaction between rotated $\widetilde{U}(1)$ charges is significantly screened along the direction parallel to the magnetic field. Comparing the susceptibility of the CFL phase at zero magnetic field, $\chi_{CFL}=2{\tilde e}^2\mu^2/(9\pi^2\Delta_{CFL}^2)$, where $\Delta_{CFL}$ is the CFL gap \cite{Chargedgluon}, with the result for MCFL at strong field, we see that the effect of the magnetic field is to decrease the electric susceptibility of the color superconductor. In Fig. 1, we plot the ratio between the parallel susceptibility of the MCFL phase and the CFL susceptibility versus the applied magnetic field (taking into account the variations with the field of $\Delta_B$ and $\Delta_M$ from \cite{SpinoneCFL}). We can see that for an acceptable value of the chemical potential ($\mu=500$ MeV), $\chi_{MCFL}^{\|}<\chi_{CFL}$ in the strong field region and that the electric susceptibility $\chi_{MCFL}^{\|}$ decreases as the magnetic field increases. This can be understood from the fact that the coherent length of the Cooper pairs, $\sim 1/\Delta_{0}$  plays the role of the dipole separation length (i.e. the  length separating the opposite charges that form the pair). The larger the field, the smaller the coherence length, so the dipole moment of the pair weakens with the increasing field since $\Delta_0$ is enhanced by the external field \cite{SpinoneCFL} in the strong field region. Therefore, the medium becomes less polarizable in the presence of a strong magnetic field.

\begin{figure}
\includegraphics[width=0.5\textwidth]{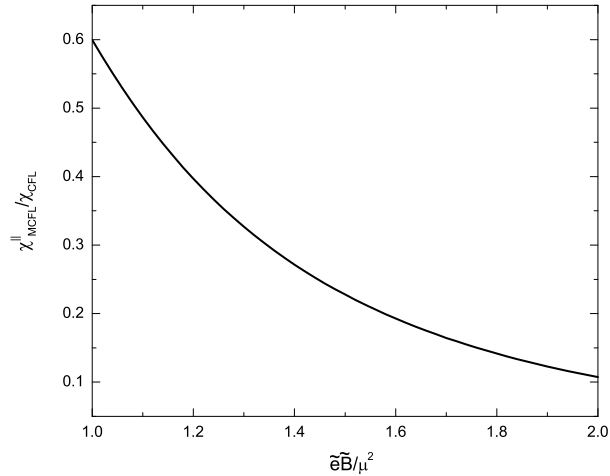}
\caption{\label{fig:wide}The ratio between the susceptibilities of the MCFL phase to that of the CFL phase ($\chi_{MCFL}^\|/ \chi_{CFL}$) vs magnetic field, for $\mu=500$ MeV. The numerical value for $\Delta_0$ and $\Delta_{CFL}$ are taken from \cite{SpinoneCFL}. The corresponding magnetic field for ${\tilde e}{\tilde B}/\mu^2=1$ is $4\times 10^{19}$ G.}
\label{drchargestrong}
\end{figure}

\section{Conclusions}

In this paper we have proven that the electric susceptibility  of a three-flavor color superconductor can be significantly modified by an applied magnetic field. Magnetic-field dependent electric susceptibilities have been obtained over the years in many condensed matter systems, where the phenomenon is known as the ME effect. The present realization of the ME effect in the context of high-dense quark matter is connected to the existence of Cooper pairs of opposite charged quarks that behave as electric dipoles with respect to the rotated electromagnetism of the MCFL phase. The pair's coherence length $\xi$ plays the role of the dipole length. When the magnetic field increases in the strong-field region, the susceptibility  becomes smaller, because the coherence length $\xi \sim 1/\Delta_0 $ decreases (i.e. $\Delta_0$ increases) with the field at a quicker rate than $\sqrt{eB}$ \cite{SpinoneCFL}. Hence with increasing magnetic field the polarization effects weaken. Physically this is easy to understand because the smaller the dipole, the more it resembles a neutral particle unable to screen an electric field. Similar arguments should be applicable in other spin-0 CS phases that leaves an in-medium electromagnetic group unbroken. Notice that the Debye screening is absent in both, the MCFL and the CFL phases, since in these phases all the quarks are paired and the pairs, being neutral with respect to the rotated electromagnetism, cannot produce the single particle monopole effect responsible for the Debye screening.

An important outcome of the present paper is the large anisotropy induced in the electric susceptibility by the strong magnetic field. This anisotropy is a direct consequence of the dimensional reduction (from 3+1 to 1+1 dimensions) that always occurs in the dynamics of charged fermions in a strong magnetic field. As a consequence, all the dipole polarization occurs only along the direction of the applied magnetic field and consequently, in the transverse direction the susceptibility is just the same as in vacuum, thus zero. In contrast, in the absence of a magnetic field, the dipoles are not oriented along any particular direction, hence, once an electric field is applied in an arbitrary direction, it will induce a medium polarization along that direction \cite{Chargedgluon}.

The anisotropic susceptibility, together with the lack of Debye screening in the MCFL phase, implies that transverse electric fields are not modified at all due to medium effects in this type of color superconductor. Presumably, the situation would be slightly different in the 2SC superconductor in a magnetic field. Even though we expect a similar anisotropy to be present in the 2SC case too, the charged blue quarks can in principle Debye screen an external electric field in all directions.

One can compare the parallel electric susceptibility of the MCFL phase,
\begin{equation}\label{susceptibility-CS}
   \chi_{MCFL}^\parallel = \frac{{2\tilde \alpha}|{\tilde e}{\tilde B}|}{3\pi\Delta_{0}^2},
\end{equation}
with the one found at zero density and strong magnetic field in massive QED \cite{Shabad},
\begin{equation}\label{susceptibility-QED}
   \chi_{QED}^\parallel = \frac{\alpha |eB|}{3\pi m^2},
\end{equation}
The two expressions are similar, with the dipole length being given by the inverse of the quasiparticle gap $\Delta_{0}$ in (\ref{susceptibility-CS}) and the inverse of the electron mass $m$ in (\ref{susceptibility-QED}). However, there is a factor 2 in (\ref{susceptibility-CS}) that is not present in (\ref{susceptibility-QED}). The factor 2 in the color superconductor comes from the contribution of two types of electric dipoles that exist in the MCFL phase associated with the two types of Cooper pairs formed by charged quarks: $\langle s_ru_b\rangle$ and $\langle d_ru_g\rangle$. In QED, on the other hand, the dipole polarization effects can only come from a single type of dipole, the one formed by the electron and the positron. Another important difference is that the polarization in the MCFL superconductor is a purely Fermi surface effect, so it requires a finite density and the formation of Cooper pairs at the Fermi surface. On the other hand, in QED the polarization is a vacuum effect, because the dipole is formed by electrons and positrons excited at the surface of the Dirac sea.

In the case of unpaired quark matter under a strong magnetic field, the parallel electric susceptibility for each quark species is similar to that of QED (\ref{susceptibility-QED}). Hence, considering all the contributions of the quarks with different colors and flavors we have
\begin{equation}\label{susceptibility-QCD}
   \chi_{QCD}^\parallel = \frac{N_c}{3\pi}\sum_{q=1}^{N_f}\alpha_q\frac{|e_qB|}{m_q^2},
\end{equation}
where $\alpha_q$, $e_q$ and $m_q$ denote the fine-structure constant, the electric charge, and the current mass of the q-th quark respectively. $N_c$ and $N_f$ are the color and flavor numbers respectively. The result (\ref{susceptibility-QCD}) is however academic, as unpaired quark matter can only exist at high energy scales, where asymptotic freedom ensures weak coupling, but (\ref{susceptibility-QCD}) was found assuming zero temperature and density.

For nuclear matter under a strong magnetic field, the main contribution to the electric susceptibility comes from the proton loop. Considering the Serot-Walecka effective field theory \cite{Serot}, and neglecting the proton magnetic moment, the electric susceptibility at strong field in nuclear matter is also very similar to that of QED with the replacement of the electron mass by the proton effective mass $\widehat{m}_P$,
\begin{equation}\label{susceptibility-NM}
   \chi_{NM}^\parallel = \frac{\alpha |eB|}{3\pi \widehat{m}_{P}^2}, \quad \widehat{m}_P=m_P-g_\sigma \overline{\sigma}
\end{equation}
In (\ref{susceptibility-NM}) $m_p$ is the bare proton mass and $g_\sigma\overline{\sigma}$ is the contribution to the proton mass of the sigma-meson mean field. At saturation density ($\rho_0=0.153 fm^{-3}$) the proton effective mass is $\sim 732 MeV$ \cite{Pacheco}. This implies that (\ref{susceptibility-NM}) is consistent only for field magnitudes $|eB|\geq 10^{20}$ G. Since the MCFL gap mass is $\Delta_0\lesssim 50$ MeV for those field strengths \cite{SpinoneCFL}, it is clear that the electric susceptibility for the magnetized color superconductor is two orders of magnitude larger than for magnetized nuclear matter. This result can lead to observable differences between neutron stars with nuclear matter cores and color superconducting cores, as discussed below.

An anisotropic susceptibility also appears in massless QED with chiral condensate (i.e. in the phenomenon of magnetic catalysis of chiral symmetry breaking MC$\chi$SB) in a strong magnetic field at zero density. In this case, chiral symmetry is spontaneously broken via the mechanism of magnetic catalysis and the parallel susceptibility is given by \cite{chiral-cond}
\begin{equation}\label{Susc-MC}
\chi_{MC\chi SB}^\parallel=\frac{\alpha |eB|}{3\pi (E^{0})^2}=\frac{\alpha}{6\pi}\exp \sqrt{\frac{4\pi}{\alpha}},
\end{equation}
with $E^0$ the LLL rest energy induced by the chiral condensation \cite{magmoment,chiral-cond}. In contrast to the previous cases (\ref{susceptibility-CS})-(\ref{susceptibility-NM}), the susceptibility (\ref{Susc-MC}) is independent of the magnitude of the applied magnetic field. The highly nonperturbative dependence on the coupling $\alpha$ makes the susceptibility (\ref{Susc-MC}), and thus the polarization in the field direction, much larger in this system than in the others previously analyzed.

Our results could be of interest for the astrophysics of compact objects. As discussed in the Introduction, the core of neutron stars could in principle reach densities and magnetic fields large enough to create the conditions needed for producing the MCFL's anisotropic susceptibility found in this paper. Under these circumstances, the anisotropic susceptibility might lead to a signature in the properties of the $\gamma$-ray bursts for stars with a color superconducting core. According to some current models \cite{Gamma} of $\gamma$-ray bursts, the violent release of energy inside the neutron star produces oscillation modes that lead to the generation of strong electric fields aligned with the star magnetic field. Electric fields generated in one magnetic pole of the star that manage to penetrate the star crust and reach the core, will be highly attenuated due to the large susceptibility in the direction parallel to the magnetic field. Then, electric fields generated in one of the two poles (along the magnetic field direction) will decay within the core and not reach the opposite pole with enough strength to produce $\gamma$-ray bursts in the other side of the star. The consequence of this effect will be very anisotropic $\gamma$-ray bursts even along the direction parallel to the magnetic field of the star. For the largest fields expected in the core of the neutron stars, this effect could be relevant only if the star's core is in either the MCFL or another similar magnetic color superconducting phase. On the other hand, for nuclear matter cores the electric susceptibility will be much smaller, as discussed above, and hence the effect will not be relevant.  Therefore, a significant anisotropy of the $\gamma$-ray bursts along the star's magnetic field direction could indicate the presence of a color superconducting quark matter core.

The anisotropic susceptibility we are reporting could be also of interest for heavy-ion collision experiments. A common feature of heavy-ion collisions is the generation of very strong magnetic fields that are produced in peripheral collisions by the positively charged ions moving at almost the speed of light \cite{HIC-B}. The colliding charged ions can generate magnetic fields estimated to be of order $eB\sim2m_{\pi}^2$ ($\sim10^{18}$ G) for the top collision, $\surd s_{NN} \sim 200$ GeV, in non-central Au-Au collisions at RHIC, or even larger, $eB\sim 15m_{\pi}^2$ ($\sim10^{19}$ G), at future LHC experiments. However, these experiments produce a hot and low-density matter that is far from the QCD-phase map region where color superconductivity is favored. On the other hand, there are planned experiments at the Beam-Energy Scan Program at RHIC, the Facility for Antiproton and Ion Research (FAIR) at GSI, the Nuclotron-based Ion Collider Facility (NICA) at JINR and the Japan Proton Accelerator Research Complex (JPARC) at JAERI. These experiments have been designed to probe the intermediate-to-large density and low temperature region of QCD phase diagram, where color superconductivity could be in principle realized and thus explored \cite{SpinoneCFL,CS-HIC}. Such experiments are expected to produce very strong magnetic fields too, so there is the possibility that they could reproduce the conditions needed for the realization of the MCFL phase. On the other hand, electric fields are also created in the collisions. Even though the maximum of the electric field can be quite large, its maximum value does not overlap with that of the magnetic field in the same spatial region \cite{HIC-E}. Hence, our approximation of strong magnetic but weak electric fields will be suitable under those circumstances. One expect that the found highly anisotropic susceptibility of the MCFL and similar superconducting phases could lead to observable anisotropic effects in the production of pairs in the directions parallel and perpendicular to the magnetic field.


\begin{acknowledgments}
This work has been supported in part by DOE Nuclear Theory grant DE-SC0002179. The authors thank M. Alford, C. Manuel, K. Rajagopal, and I. Shovkovy for enlightening discussions and comments.

\end{acknowledgments}

%


\begin{thebibliography}{99}


\bibitem{CS}M. G. Alford, A. Schmitt, K. Rajagopal and T. Sch{$\ddot a$}fer, {\itshape Rev. Mod. Phys.} 80 (2008) 1455 {\itshape and references therein}.

\bibitem{Magnetars}C. Thompson and R. C. Duncan, {\itshape Astrophys. J.} 392 (1992) L9, {\itshape ibid} 473 (1996) 322; S. Kulkarni and D. Frail, {\itshape Nature} 365 (1993) 33; T. Murakami {\itshape et al.}, {\itshape Nature} 368 (1994) 127; Ibrahim {\itshape et al.}, {\itshape Astrophys. J}. 609 (2004) L21.

\bibitem{virial}L.~Dong and S.~L.~Shapiro \textit{ApJ.} 383 (1991) 745; M. Bocquet, \textit{et. al}, \textit{Astron. Astrophys.} 301 (1995) 757.

\bibitem{EoS-H} E. J. Ferrer, \textit{et. al}, \textit{Phys. Rev. C} 82 (2010) 065802; L. Paulucci, \textit{et. al},  \textit{Phys. Rev. D} 83 (2011) 043009.

\bibitem{CFL}M. Alford, K. Rajagopal and F. Wilczek, {\itshape Nucl. Phys. B} 537 (1999) 443.

\bibitem{ABR}M. Alford, J. Berges and K. Rajagopal, {\itshape Nucl. Phys. B} 571 (2000) 269; E. V. Gorbar, \textit{Phys. Rev. D} 62 (2000) 014007.

\bibitem{2SC} D. Bailin and A. Love, \textit{Phys. Rept.} 107 (1984) 325; M. Alford, K. Rajagopal and F. Wilczek, \textit{Phys. Lett. B} 422 (1998) 247; R. Rapp, \textit{et. al}, \textit{Phys. Rev. Lett.} 81 (1998) 53.

\bibitem{MCFL}E. J. Ferrer, V. de la Incera and C. Manuel, {\itshape Phys. Rev. lett.} 95 (1005) 152002; {\itshape Nucl. Phys. B} 747 (2006) 88; {\itshape PoS JHW2005} (2006) 022; {\itshape J. Phys. A} 39 (2006) 6349.

\bibitem{MCFLoscillation}J. L. Noronha and I. A. Shovkovy, {\itshape Phys. Rev. D}76 (2007) 105030; K. Fukushima and H. J. Warringa, {\itshape Phys. Rev. Lett.} 100 (2008) 03200.

\bibitem{phases}E. J. Ferrer and V. de la Incera, {\itshape Phys. Rev. D} 76 (2007) 045011; {\itshape AIP Conf. Proc.} 947 (2007) 395.


\bibitem{SpinoneCFL}B. Feng, E. J. Ferrer and V. de la Incera, \textit{Nucl. Phys. B} 853 (2011) 213.

\bibitem{magmoment}E. J. Ferrer and V. de la Incera, \textit{Phys. Rev. Lett.} 102 (2009) 050402; \textit{Nucl. Phys. B} 824 (2010) 217.


\bibitem{QCDcondense}K. Rajagopal and F. Wilczek, {\itshape The Condensed Matter Physics of QCD} "At the Frontier of Particle Physics/Handbook of QCD", M. Shifman, ed., (World Scientific, Singapore, 2001). v2, arXiv:hep-ph/0011333.

\bibitem{early-ME}W. C. Rontgen, {\itshape Ann. Phys.} 35 (1888) 264; H. A. Wilson, {\itshape Phil. Trans. R. Soc. A} 204 (1905) 129.

\bibitem{Curie}P. Curie, {\itshape J. Physique} 3 (1894) 393.

\bibitem{L-L}L. D. Landau and E. M. Lifshitz, "{\itshape Electrodynamics of Continuous Media}" 1960 (Oxford: Pergamon).

\bibitem{Revival}For a review on the topic see M. Fiebig, {\itshape J. Phys. D: Appl. Phys.} 38 (2005) R123.

\bibitem{Anderson}P. W. Anderson, {\itshape Phys. Rev. Lett.} 3 (1959) 325.

\bibitem{Abrikosov} A. A. Abrikosov and L. P. Gorkov, \textit{Zh. Eksp. Teor. Fiz.} 39 (1960) 1781 [\textit{Sov. Phys. -JETP} 12 (1961) 1243]; P. G. de Genness , \textit{Superconductivity of Metals and Alloys} (W. A. Benjamin Inc, New York, 1966).

\bibitem{ferroelectricity} W. Känzig, \textit{Ferroelectrics and Antiferroelectrics} in F. Seitz, T. P. Das, D. Turnbull, and E. L. Hahn, Solid State Physics, 4 (Academic Press. p. 5, 1957); M. Lines and A. Glass, \textit{Principles and applications of ferroelectrics and related materials}, (Clarendon Press, Oxford, 1979).

\bibitem{Ritus}V. I. Ritus, {\itshape Ann. Phys.} (N.Y.) 69 (1972) 555; {\itshape Zh. Eksp. Teor. Fiz.} 75 (1978) 1560 [{\itshape Sov. Phys. JETP} 48 (1978) 788]; E. Elizalde, E. J. Ferrer and V. de la Incera, {\itshape Ann. Phys.} (N.Y.) 295 (2002) 33; {\itshape Phys. Rev. D} 70 (2004) 043012.


\bibitem{Ng}C. N. Leung, Y. J. Ng and A. W. Ackley, {\itshape Phys. Rev. D} 54 (1996) 4181; D. -S. Lee, C. N. Leung and Y. J. Ng, {\itshape Phys. Rev. D} 55 (1997) 10.


\bibitem{Chargedgluon}D. F. Litim and C. Manuel, {\itshape Phys. Rev. D} 64 (2001) 094013.

\bibitem{Rischke}A. Schmitt, Q. Wang and D. H. Rischke, {\itshape Phys. Rev. D} 69 (2004) 094017.

\bibitem{Shabad}
I. A. Batalin and A. E. Shabad, Zh. Eksp. Teor. Fiz. 60 (1971) 894 [\textit{Sov. Phys.- JETP} 33 (1971) 483]; A. E. Shabad, \textit{Lettere al Nuovo Cim.} 2 (1972) 457; Wu-Y. Tsai \textit{Phys. Rev. D} 10 (1974) 2699; V. N. Baier, V. M. Katkov and V. M. Strakhovenko, \textit{ZhETF} 68 (1975) 403; A. E. Shabad, \textit{Ann. Phys. (N.Y.)} 90 (1975) 166; A. E. Shabad, \textit{Sov. Phys., Lebedev Inst. Rep.} 3 (1976) 11; D. B. Melrose and R.J. Stoneham, \textit{Nuovo Cim.} 32 (1977) 435; A. E. Shabad and V. V. Usov, \textit{Phys. Rev. D} 77 (2008) 025001.


\bibitem{Serot} B.D. Serot, \textit{Phys. Lett. B} 86 (1979) 146; B.D. Serot and J.D. Walecka, \textit{Adv. Nucl. Phys.} 16 (1986) 1-327; \textit{Int. J. Mod. Phys. E} 6 (1997) 515; J. Boguta and A. Bodmer, \textit{Nucl. Phys. A} 292 (1977) 413; J. Zymanyi and S.A. Moszkowski, \textit{Phys. Rev. C} 42 (1990) 1416; C. Fuchs, H. Lenske and H.H. Wolter, \textit{Phys. Rev. C} 52 (1995) 3043; S. Typel and H.H. Wolter, \textit{Nucl. Phys. A} 656 (1999) 331; G. Hua, L. Bo and M. Di Toro, \textit{Phys. Rev. C} 62 (2000) 035203.

\bibitem{Pacheco} G.F. Marranghello, C.A.Z. Vasconcellos and J.A. de Freitas Pacheco, {\itshape Phys. Rev. D} 66 (2002) 064027; G.F. Marranghello, C.A.Z. Vasconcellos and J.A. de Freitas Pacheco and M. Dllig, \textit{Int. J. Mod. Phys. E} 11 (2002) 83.


\bibitem{chiral-cond} E. J. Ferrer, V. de la Incera and A. Sanchez, {\itshape Phys. Rev. Lett.} 107 (2011) 041602.

\bibitem{Gamma} P.A. Sturrock, Nature 321 (1986) 47; I.A. Smith and R.I. Epstein, \textit{Relativistic Electron-Positron Beams from Oscillating Neutron Stars,"} in Proceedings of the Los Alamos Workshop on Gamma-Ray Bursts, Edit. by Cheng Ho, R.I. Epstein and E.E. Fenimore, Cambridge, Univ. Press, 1992, p. 29.

\bibitem{HIC-B}D.E. Kharzaeev, L.D. McLerran and H.J. Warringa, \textit{Nucl. Phys. A} 803 (2008) 227; V.V. Shokov, A. Yu. Illarionov and V.D. Toneev, \textit{Int.  J. Mod. Phys. A}  24 (2009) 5925.

\bibitem{CS-HIC}D.B. Blaschke, \textit{et. al}, \textit{Acta Phys. Polon. Supp.} 3 (2010) 741; T. Kl$\ddot{a}$hn, D. Blaschke and F. Weber, arXiv: 1101.6061 [nucl-th].

\bibitem{HIC-E}V. Voronyuk, et al., \textit{Phys. Rev. C} 83 (2011) 054911.




\end{thebibliography}
\end{document}